\documentclass[12pt,preprint]{aastex}

\shorttitle{Kinematic and Spectral Ages of CTD~93}
\shortauthors{Nagai, H. et al.}

\begin{document}


\title{THE KINEMATIC AND SPECTRAL AGES OF THE COMPACT RADIO SOURCE CTD~93}


\author{Hiroshi NAGAI\altaffilmark{1,2}}
\email{nagai@hotaka.mtk.nao.ac.jp}

\author{Makoto INOUE\altaffilmark{2}}

\author{Keiichi ASADA\altaffilmark{2}}

\author{Seiji KAMENO\altaffilmark{2, *}}

\and

\author{Akihiro DOI\altaffilmark{3}}


\altaffiltext{1}{Department of Astronomical Science, 
The Graduate University for Advanced Studies, Osawa 2-21-1, 
Mitaka, Tokyo 181-8588, Japan}
\altaffiltext{2}{National Astronomical Observatory of Japan, 
Osawa 2-21-1, Mitaka, Tokyo 181-8588, Japan}
\altaffiltext{3}{Department of Physics, Faculty of Science, 
Yamaguchi University, Yoshida, Yamaguchi 753-8512}
\altaffiltext{*}{Present address: Department of Physics, Faculty of Science, Kagoshima University, Korimoto 1-21-35, Kagoshima 890-0065, Japan}


\begin{abstract}
We present a study of the kinematic and spectral ages of the Gigahertz-Peaked Spectrum (GPS) source CTD~93.  Measurements of the hot spot separation over 8.5\,yr show evidence of an increase.  The separation rate along the source axis is 0.34$\pm$0.11$c$ (H$_{0}$=72\,km s$^{-1}$ Mpc$^{-1}$), which results in a kinematic age of 2200$\pm$700~yr.  Assuming that two hot spots are moving apart at equal speeds, we derive an advance speed of 0.17$\pm$0.06$c$.  The radio lobe spectra show a high frequency steepening, as expected if energetic electrons lose energy by synchrotron radiation.  The spectral break decreases with the distance from the hot spot in the northern component of CTD~93.  This tendency is expected from the basic scenario of radio lobe evolution involving particle acceleration at the hot spots, with the radio lobes populated by high energy electrons which have leaked from the hot spots.  Although a core-jet morphology for CTD~93 has previously been proposed, these results indicate that the morphology is similar to that of Compact Symmetric Object (CSO).  From the spectral fits in the northern component we found a break frequency of 3.7\,GHz at the edge of the lobe.  The resultant spectral age is $\sim$300\,yr assuming the equipartition magnetic field.  This requires the advance speed of 0.26$c$, which shows a good agreement of the hot spot motion of 0.17$\pm$0.06$c$.  Our results strongly support the hypothesis that CSOs are young radio sources.
\end{abstract}	


\keywords{galaxies: active --- galaxies: individual(CTD~93) ---
  galaxies: jets --- galaxies: nuclei --- galaxies: evolution ---
  radio continuum: galaxies}



\section{INTRODUCTION}
Gigahertz Peaked Spectrum (GPS) sources are compact radio sources that show a convex radio spectrum with a peak between about 0.5 and 10\,GHz (O'Dea 1998).  GPS sources constitute about 10\% of powerful radio sources at centimeter wavelengths (O'Dea 1998).  Although some GPS sources have an extended structure (e.g., Stanghellini et al.\ 1998), their sizes are typically within 1\,kpc.  Early VLBI observations resolved some GPS sources into two relatively symmetric components. These sources were called Compact Doubles (CDs) (Phillips \& Mutel 1980, 1982).  Phillips \& Mutel proposed that CDs are at a young stage of their evolution to classical double radio sources, which is referred to as the youth scenario.  Alternatively, it has been suggested that CDs could be normal aged radio sources that are confined by the dense plasma and unable to escape (van Breugel et al.\ 1984), the frustration scenario.

More recent high dynamic range VLBI observations revealed that a number of CDs contain multiple components.  Such sources with a core and two-sided symmetrical radio components are named Compact Symmetric Objects (CSOs) (Wilkinson et al.\ 1994; Readhead et al.\ 1996).  CSOs are currently favored as young radio galaxies because the hot spot motions that have been detected in some of them (e.g., Polatidis et al.\ 1999; Conway 2002; Gugliucci et al.\ 2005) yield kinematic ages that are $\sim1000$ times younger than classical radio galaxies such as Cygnus~A.  An alternative means of estimating source ages is the use of spectral aging.  Murgia (2003) analyzed the total spectra of a large sample of CSS and some CSOs, and found spectral ages in the range of
$10^{2}-10^{5}$ years.

The luminous radio source CTD~93 (B~1607+268, J~1609+2641) is one of the earliest radio sources known to be a GPS source (Kraus et al.\ 1968).  Its radio morphology is classified as CD (Phillips \& Mutel 1980), and it has a spectral peak around 1\,GHz (Stanghellini et al.\ 1998).  The radio luminosity is $3\times10^{27} h^{-2}$W Hz$^{-1}$ at 1.6~GHz (Shaffer et al.\ 1999).  CTD~93 is optically identified with a galaxy of 20.3 magnitude and has a measured red-shift of 0.473 (Phillips \& Shaffer 1983; O'Dea et al.\ 1991).  This source has a quite symmetric double-lobe--like structure, and so is a candidate CSO.  However, high dynamic range observations by Shaffer et al.\ (1999) did not find any feature between the two dominant radio components that might correspond to the core. They claimed that the northern component consists of the core and jet, while the southern component is the reappearance of the jet.  Alternatively, the absence of the core candidate could be explained by synchrotron self-absorption (SSA) and/or Free-Free absorption (FFA).  Another possible explanation is that the activity of the central component has stopped.

In this paper we derive the kinematic and spectral ages of CTD~93.  Because CTD~93 has two predominant radio components, it is possible to estimate the kinematic age by measuring the positional change of one component relative to the other.  This is an important study to judge whether GPSs are young or frustrated.  Besides, estimation of the spectral age by measuring the spectral break is an indicator of the source age (e.g., Myers \& Spangler 1985; Carilli et al.\ 1991; Murgia et al.\ 2003).  It is also interesting to investigate the relation between the kinematic age and the spectral age.

\section{SPECTRAL AGING}\label{sec2}
The basic theory of the synchrotron aging process in radio sources was developed by Kardashev (1962) and Pacholczyk (1970), often referred to as the KP model, and by Jaffe \& Perola (1970), called the JP model.  The KP and JP models assume a single impulsive injection to produce the power-law distribution of relativistic electrons.  The KP model assumes that all electrons maintain the same pitch angle during their radiative lifetimes.  On the other hand, the JP model allows the pitch angle scattering, and the scattering time scale is assumed much shorter than the radiative lifetime.  A third model, the CI model, assumes that continuous injection of relativistic electrons over the lifetime of the source (Kardashev 1962; Pacholczyk 1970).  Thus the CI model can be adopted for regions where it is hard to distinguish aged electrons from newly injected electrons.  Therefore, if we are considering resolved sources, the KP and JP models are adequate.  Indeed, Murgia (2003) reported that while the total spectra of CSS sources are well fitted to the CI model, the spectra in the spatially resolved cases are fitted to the JP model.  Thus we can omit the CI model in this paper.

For the simplicity, we assume that the distribution of electrons is only affected by synchrotron radiation losses.  Non-thermal electrons in the shock regions such as hot spots usually form a power-law distribution, $N(E)=N_{0}E^{-\gamma}$, where $N_{0}$ is the electron density at $t=0$.  The distribution of electrons at some later time $t$ is given by
\begin{equation}
N(E, \theta, t)=N_{0}E^{-\gamma}(1-c_{2}H^{2}\sin^{2}{\theta}Et)^{\gamma-2},
\end{equation}
where $H$ is the magnetic field strength, and $\theta$ is the pitch angle between the electron and magnetic field.  The quantity $c_{2}=2e^{4}/3m^{4}c^{7}$ is a constant defined by Pacholczyk (1970), where $e$ is the electron charge, $m$ is the electron mass, and $c$ is the light velocity.
  
In the KP model, it is assumed that the pitch angle distribution of electrons is isotropic, and the electron density and magnetic field are homogeneous along the source depth $s$.  Then the radiation spectrum in an optically thin medium is given by
\begin{eqnarray}
S_{\nu}(t)&=& s\int_{4\pi}c_{3}H\sin{\theta}\int_{0}^{E_{H}}F(x) N_{0}E^{-\gamma}(1-c_{2}H^{2}\sin^{2}{\theta}Et)^{\gamma-2} dEd\Omega \nonumber \\
&=&S_{0} \nu^{\frac{1-\gamma}{2}}\int_{0}^{\frac{\pi}{2}}d\theta \int_{\frac{\nu}{\nu_{b}}\sin^{3}{\theta}}^{\infty}dx\sin^{\frac{3\gamma+1}{2}} {\theta} F(x) x^{-\frac{1}{2}} (x^{\frac{1}{2}}-(\frac{\nu}{\nu_{b}}\sin^{3}{\theta})^{1/2})^{\gamma-2},
\end{eqnarray}
where $S_{0}$ is a constant factor, $\nu_{b}=c_{1}/c_{2}^{2}H^{3}t^{2}$, $F(x)=x\int_{x}^{\infty}K_{5/3}(z)dz$, and $x$ is the frequency normalized by the critical frequency (Pacholczyk 1970).  The parameters $c_{1}=3e/4\pi m^{3}c^{5}$ and $c_{3}=\sqrt{3}e^{3}/4\pi mc^{2}$ are constant factors defined by Pacholczyk (1970).  The resultant spectrum is steeper above the break frequency $\nu_{b}$ than the initial injection spectrum.  The break frequency can be expressed in a more convenient way as
\begin{equation}
\frac{\nu_{b}}{\mathrm{[GHz]}}=1.12\times10^{9} (\frac{H}{\mathrm{[mG]}})^{-3} (\frac{t}{\mathrm{[yr]}})^{-2}.
\end{equation}

In case of the JP model, the terms of $c_{2}H^{2}\sin^{2}{\theta}$ should be replaced by $c_{2}H^{2}<\sin^{2}{\theta}>$, where $<>$ represents the ensemble average.  Thus the radiation spectrum is given by
\begin{equation}
S_{\nu}=S_{0}\nu^{\frac{1-\gamma}{2}}\int_{0}^{\frac{\pi}{2}}d\theta \int_{\frac{\nu}{\nu_{b}\sin{\theta}}}^{\infty}dx \sin^{\frac{\gamma+3}{2}}{\theta}F(x)x^{-\frac{1}{2}}(x^{\frac{1}{2}}-(\frac{\nu}{\nu_{b}\sin{\theta}})^{1/2})^{\gamma-2},
\end{equation}
and the break frequency is given by  
\begin{eqnarray}
\nu_{b}&=&\frac{9}{4}\frac{c_{1}}{c_{2}^{2}}\frac{1}{H^{3}t^{2}} \nonumber \\
\frac{\nu_{b}}{\mathrm{[GHz]}}&=&2.52\times10^{9} (\frac{H}{\mathrm{[mG]}})^{-3} (\frac{t}{\mathrm{[yr]}})^{-2}.
\end{eqnarray}
This results in a spectrum that falls exponentially above the break frequency. 

Although the shape of the emission spectrum depends on the model, both the KP and JP models lead to a high frequency steepening above the break frequency.  The break frequency moves to lower frequency with time.

As we have seen above, both KP and JP models have a number of assumptions.  Realistic model must include other physical effects.  Possible effects that change the spectral shape are inhomogeneous magnetic field and electron density, time variation of magnetic field, confusion from multiple particle populations, and so on (e.g., Rudnick 2002; Kaiser 2005).  These effects make the spectral age estimates difficult.  Calculation of these effects is beyond the scope of this paper.
\section{DATA ANALYSIS}	
\subsection{Observation and Data Reduction}\label{reduction}
\begin{figure}[H]
\begin{center}
\plotone{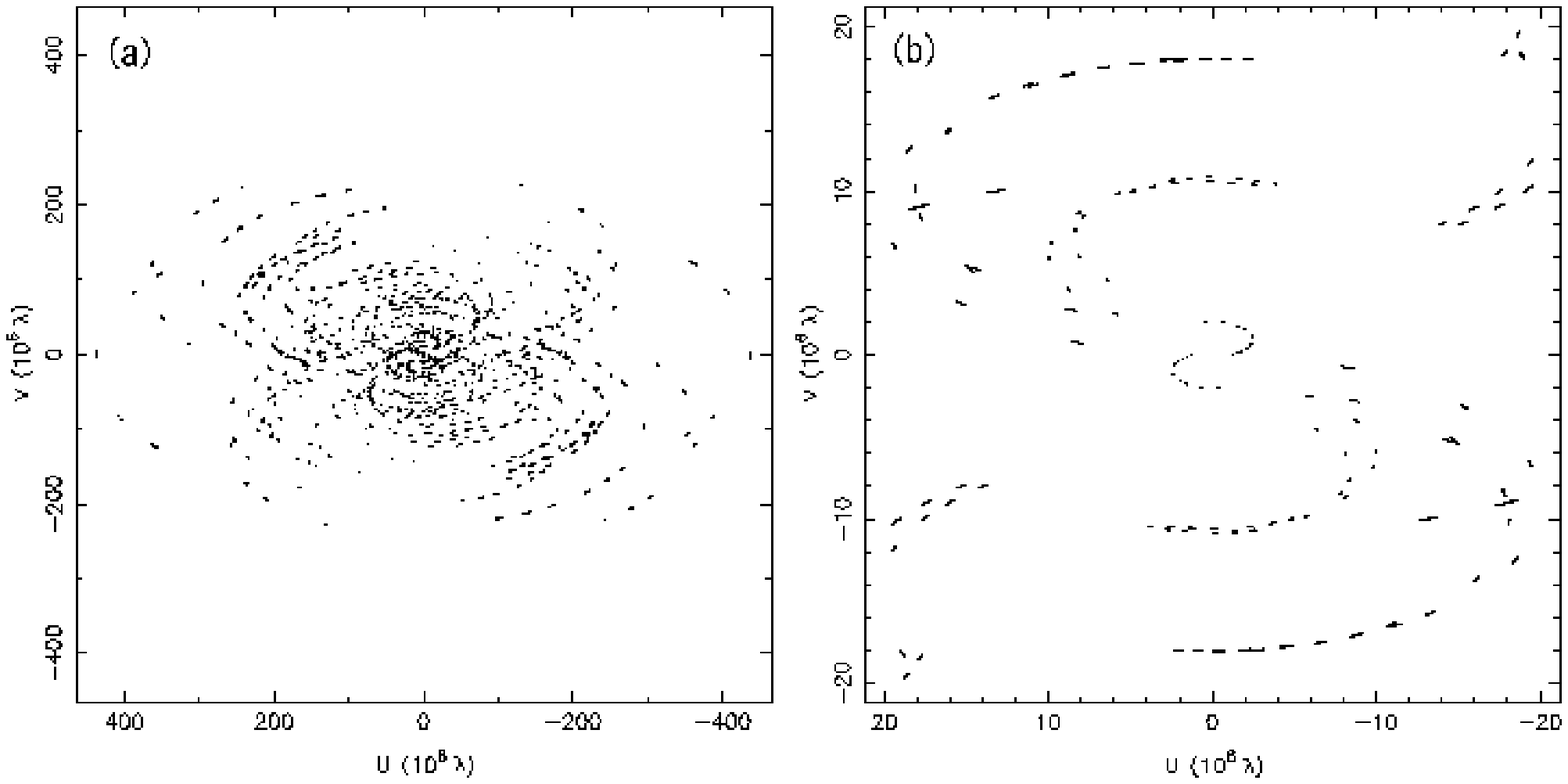}
\caption{(a)The ($u,v$) coverage at 15.285~GHz. (b)The ($u,v$) coverage of the inner 20~M$\lambda$ wavelength at 15.285 GHz.}
\label{uvplot}
\end{center}
\end{figure}

The observation was carried out using ten VLBA stations at 2.244, 2.364 (13\,cm), 4.99 (6\,cm), 8.413 (4\,cm) and 15.285\,GHz (2\,cm) on 2003 October 25.  In addition, a single VLA antenna was used at 4.99, 8.413, 15.285~GHz to add sensitivity to any extended structure.  Figure~\ref{uvplot} shows the visibilities in the ($u,v$) plane.  We performed a number of snap shots over a wide range hour angle in order to obtain good ($u,v$) coverage at all frequency bands.  Scans of 7~minutes were used at each frequency band except for 15.285~GHz, where 11~minutes scans were used.  The total integration time consists of 11 scans at all frequency bands.  Four base band converters (BBCs) were used at 4.99, 8.415, and 15.285~GHz, while two BBCs were used at 2.244 and 2.364~GHz.

We performed the data reduction using the Astronomical Image Processing System (AIPS) software package developed by the National Radio Astronomy Observatory (NRAO).  Amplitude calibration for each antenna was derived using measurements of the system temperatures during the observation.  Fringe fitting was performed with the AIPS task FRING.  The fringe solutions of the baselines including MK and SC antennas were not determined at some time ranges at 15.285~GHz.  After delay and rate solutions had been applied, the data were averaged across bandwidths of 16\,MHz at 2.244 and 2.364~GHz, and 32\,MHz at other frequency bands.  Imaging was performed using CLEAN and self-calibration in the Difmap software package (Pearson, Shepherd, \& Taylor~1994).

In the spectral aging analysis, we need to carefully consider resolution effects.  The lack of short baselines could result in flux being missed for extended structures.  In multi-frequency observations with an identical array configuration, the missing flux is more serious at higher frequencies.  Such observations could show a steeper spectrum at higher frequencies, mimicking the effects of synchrotron losses.  The PT antenna and VLA antenna provide the shortest baseline length in the ($u,v$) coverage in our observation.  At 15.285~GHz, the minimum ($u,v$) radius $\sqrt{u^{2}+v^{2}}$ is about 1.2~M$\lambda$ wavelengths.  This baseline length results in a missing flux of only a few percent for an extended structure of $\sim$20~mas.  Thus it is not necessary to consider the effect of missing flux as long as we are discussing the spectral aging within 20~mas.
	
We also obtained the data sets of two archival 2\,cm observations, BS025 (Shaffer et al.\ 1999) and BR077.  BS025 was performed on 1995 July 23 at 15.285\,GHz, and BR077 on 2002 May 5 at 15.335\,GHz, both using 10 VLBA stations.  The process of data reduction is the same as described above.  Detailed observational parameters of BS025 are provided in Shaffer et al.\ (1999).  The data of BR077 contains eight 8-MHz basebands with a total integration time of 13 minutes.

\subsection{Spectral Model Fitting}\label{SpectralModelFitting}
The two spectral aging models, KP and JP, were fitted to the five frequency data.  The fitting parameters of both models are the initial injection index $\gamma$ ($\gamma=2\alpha+1$ where $\alpha$ is the emission spectrum index, defined in the sense $S \propto \nu^\alpha$), the break frequency $\nu_{b}$, and intensity scaling $S_{0}$.  Although both models have three free parameters, we fixed the value of $\gamma$.  This assumption implies no time variation of the injection rate.  The injection index is related to the shock compression ratio in the Fermi acceleration process.  In the case of a strong non-relativistic shock, $\gamma=2.0$ (Bell 1978; Blandford \& Ostriker 1978).  However, it has been reported that $\gamma$ depends on the shock conditions (Heavens \& Drury 1988; Ballard \& Heavens 1991).  Also it is not clear what is the most plausible model for acceleration in the hot spots.  A few examples which hint at the injection index for hot spots have been reported by Carilli et al.\ (1991) and Meisenheimer et al.\ (1989), who found that the hot spots in Cygnus~A and three other radio galaxies have $\gamma=2.0$.  This suggests that hot spots are not far from the condition of a strong non-relativistic shock.  Thus we have generally restricted our analysis to $\gamma=2.0$, but will mention the injection index dependence of the fitting in section \ref{InjectionDependence}.
	
In the fitting process, we first performed the fitting to the ratio of intensities at two adjacent frequencies to minimize the value of $\chi^{2}=\sum_{i=1, j=2}^{5}(S_{\nu_{j}}^{num}/S_{\nu_{i}}^{num}-S_{\nu_{j}}^{obs}/S_{\nu_{i}}^{obs})^{2}$, where $S_{\nu_{i}}^{num}$ is the numerically computed flux density value, $S_{\nu_{i}}^{obs}$ is the observed flux density, and $\nu_{i=1, 2, 3, 4, 5}=$2.224, 2.364, 4.99, 8.415, and 15.285~GHz, respectively.  This routine provides $\nu_{b}$ independent of $S_{0}$.  The $S_{\nu_{j}}^{num}/ S_{\nu_{i}}^{num}$ was calculated in increments of 0.1~GHz for $1.5\le\nu_{b}\le35$~GHz.  Then we searched for the $S_{0}$ which minimized the value of $\chi^{2}$.

\section{RESULTS}
\subsection{Total Intensity \& Model Fits}
\begin{figure}[H]	
\begin{tabular}{cc}
\begin{minipage}{1.0\hsize}
\plottwo{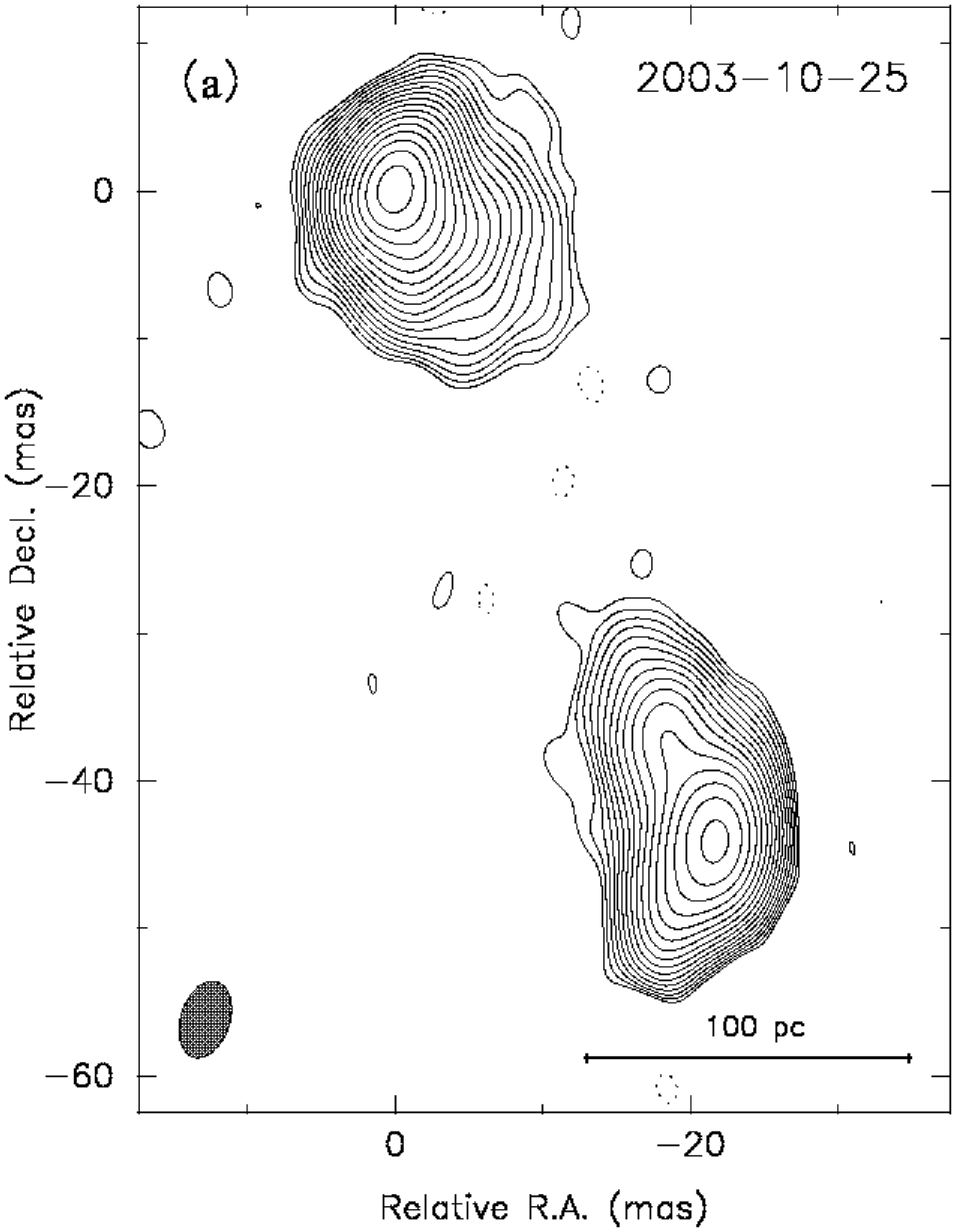}{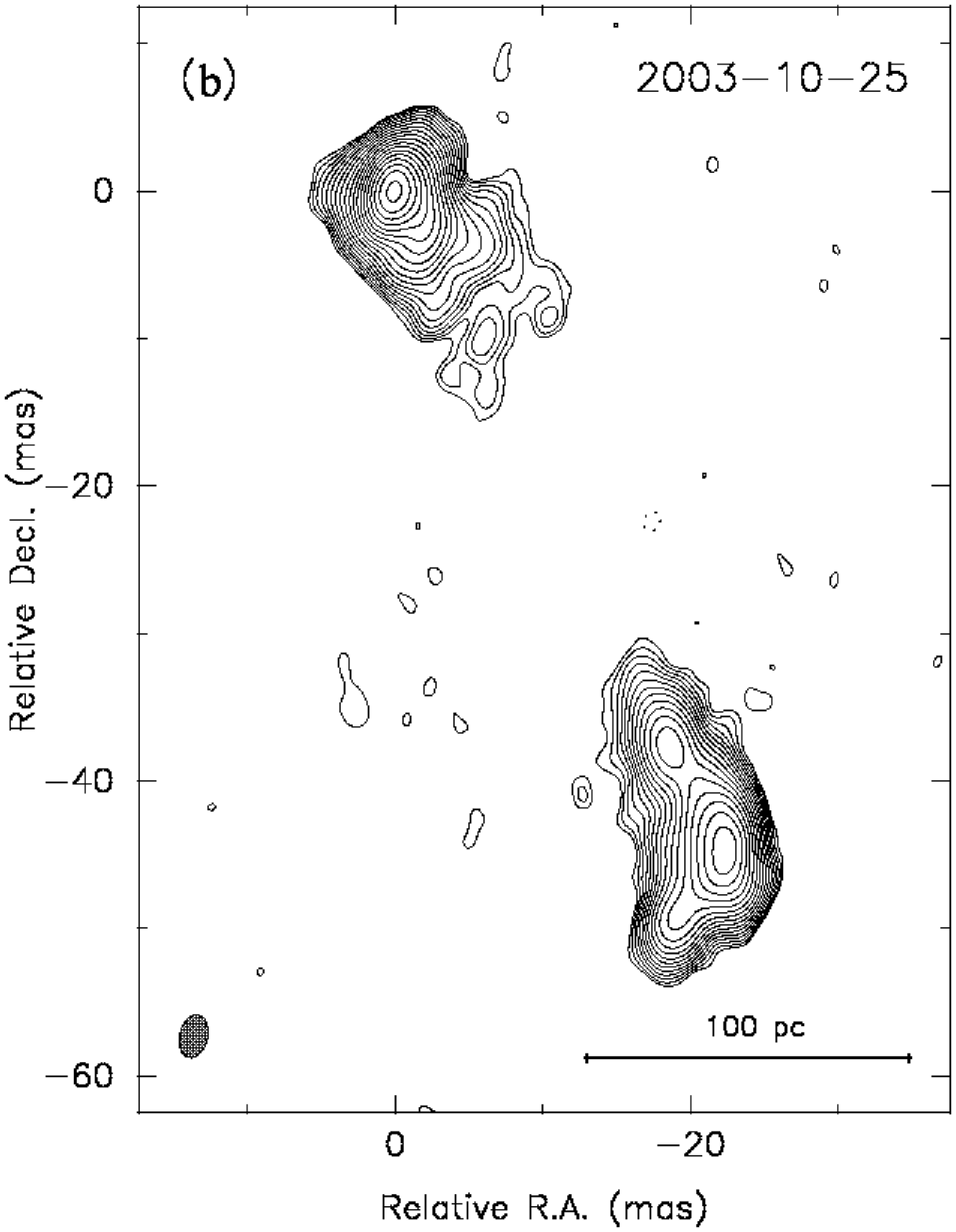}
\end{minipage}\\
\begin{minipage}{1.0\hsize}
\plottwo{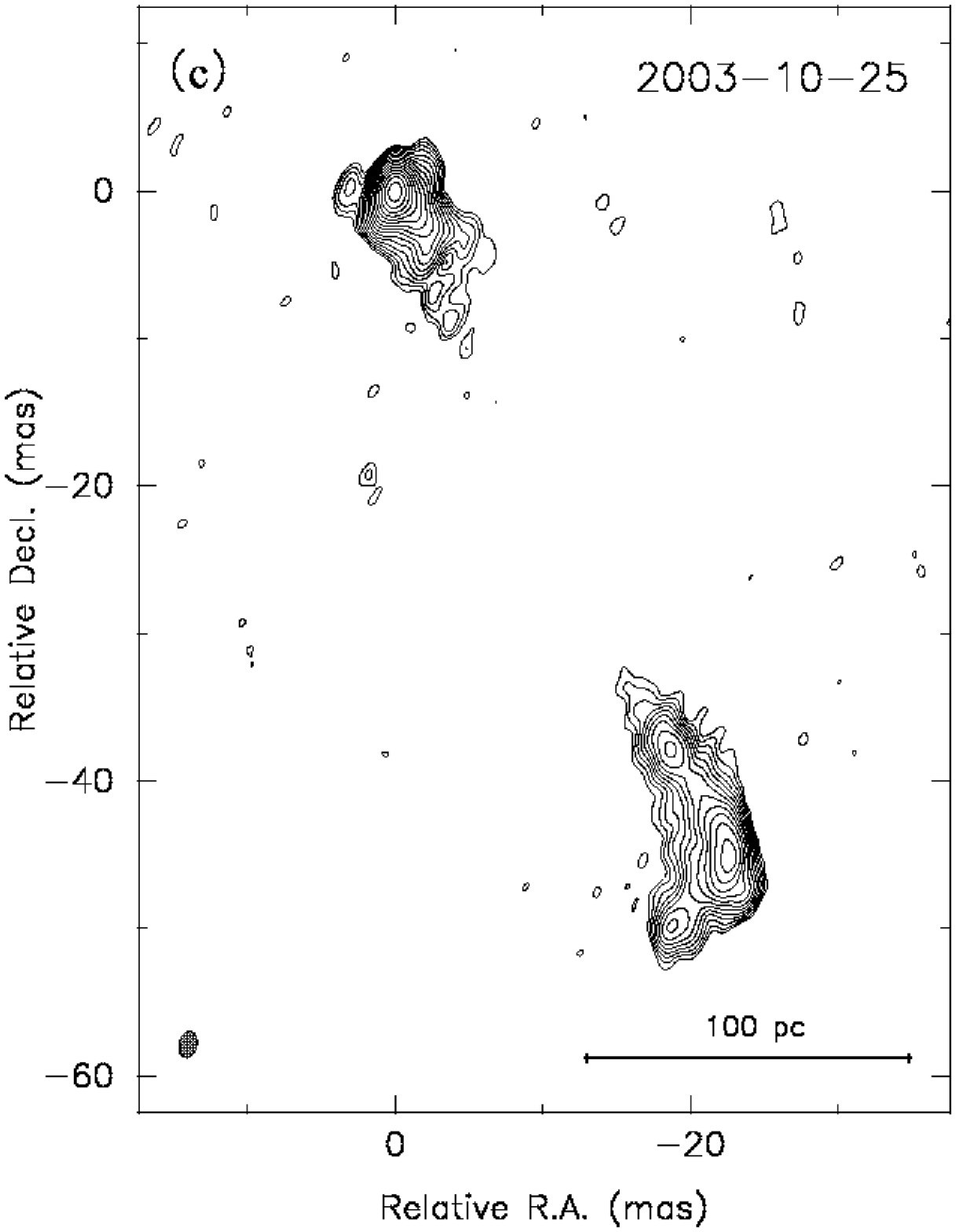}{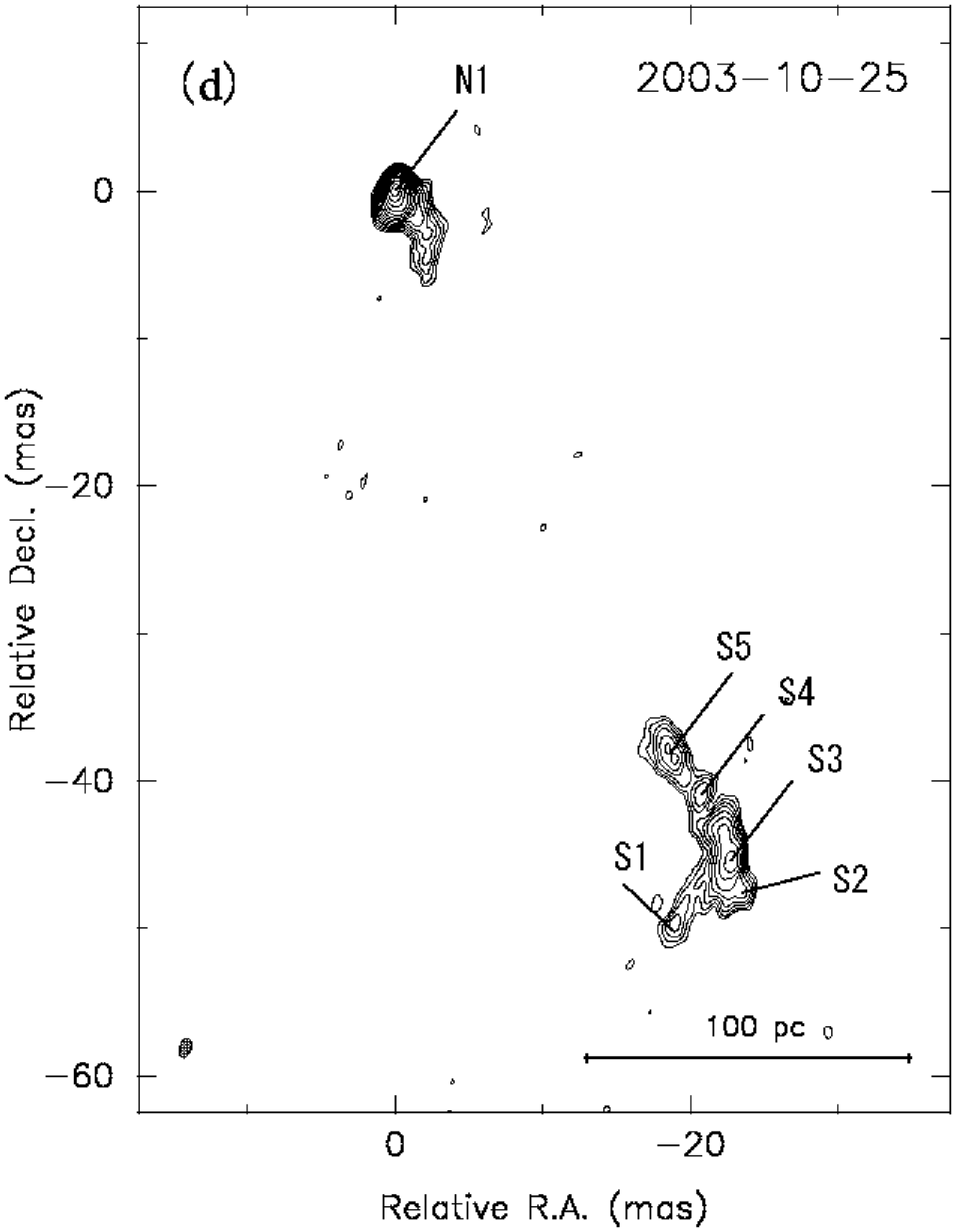}
\end{minipage}
\end{tabular}
\caption{The total intensity images at (a) 2.244\,GHz, (b) 4.99\,GHz, (c) 8.415\,GHz, and (d) 15.285\,GHz.  The contours are $\sqrt{2}n (n=$-$1, 1, 2, 4, \dots, 512) \times3.1$\,mJy, 0.69\,mJy, 0.717\,mJy, and 0.75\,mJy which are 3-times the image noise r.m.s.\ at the respective frequencies.  The beam FWHM of each image is $5.33\times3.3$\,mas at $-18^{\circ}$, $2.96 \times1.88$\,mas at $-14^{\circ}$, $1.81 \times1.17$\,mas at $-13.7^{\circ}$, and $1.34\times0.747$\,mas at $-18.9^{\circ}$, respectively.}
\label{TotalIntensity}
\end{figure}

Figures \ref{TotalIntensity}(a)-(d) show the total intensity map at
2.244, 4.99, 8.415, and 15.285~GHz, respectively.
All images show two symmetric structures along a position angle of $\sim$25~degrees.  The brightness peaks of the two components are separated by about 50~mas, which corresponds to 240~pc.  Both components show a tail of emission extended toward the center of the source.  In the 2-cm image, the southern lobe is seen as a collimated jet and is resolved into several knotty structures.

\begin{figure}[H]
\begin{center}
\plottwo{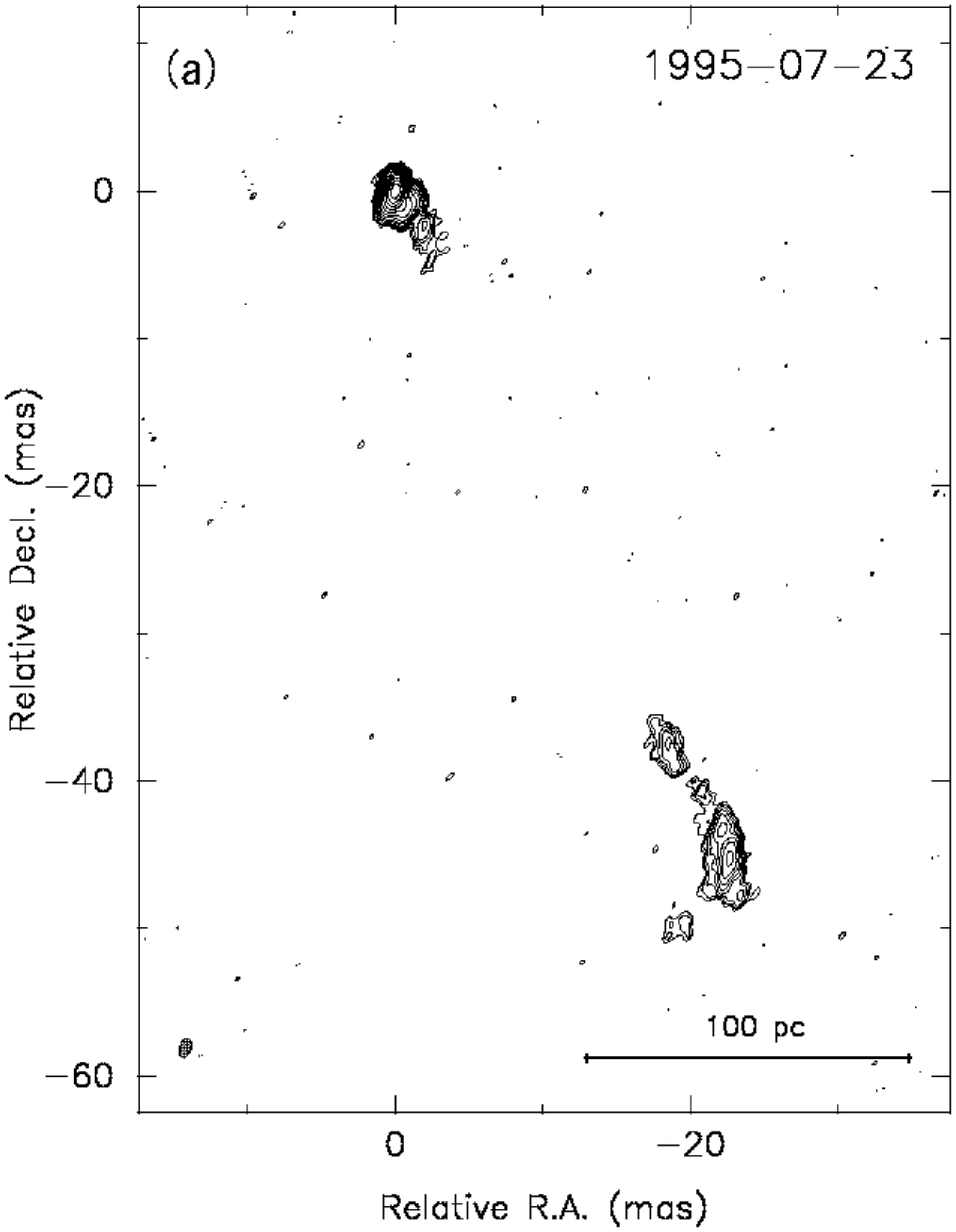}{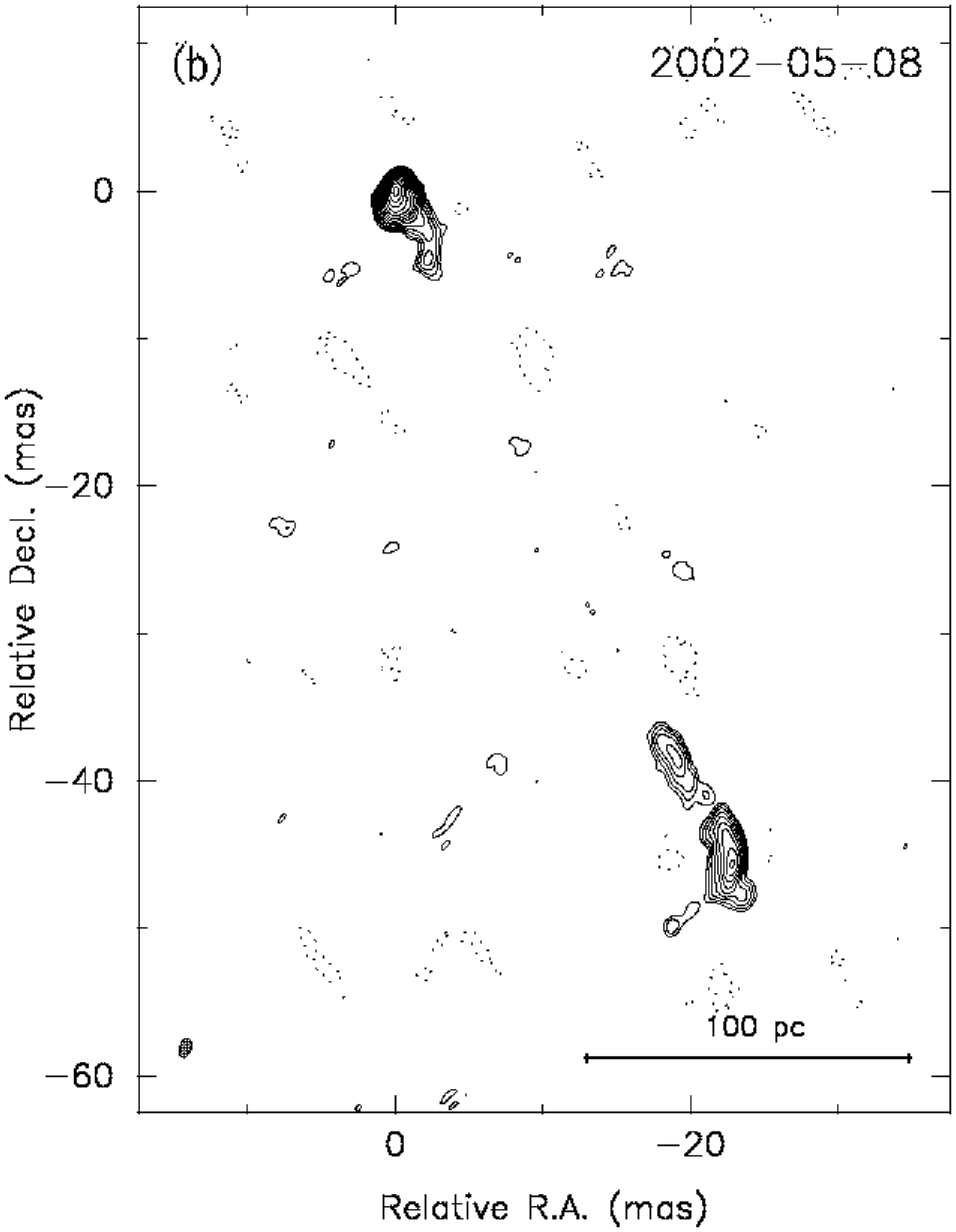}
\end{center}
\caption{(a) VLBA contour maps of BS025 at 15\,GHz and (b) BR077 at 15\,GHz.  
Both images are tapered to same resolution of our 2-cm image.  The contour levels start with 1.1 and 0.93 mJy/beam and increase from there by factors of $\sqrt{2}$.}
\label{archives}
\end{figure}

We are interested in the relative kinematics of the components of the source.  To measure the position change of components as a function of time, we compared the position of our 2-cm image with those of the previous observations BS025 and BR077.  The fully calibrated images of these data sets are shown in Figure~\ref{archives}.  They are tapered to the same resolution of our 2-cm image.

\begin{figure}[H]
\begin{tabular}{ccccc}
\begin{minipage}{0.33\hsize}
\includegraphics[width=5cm]{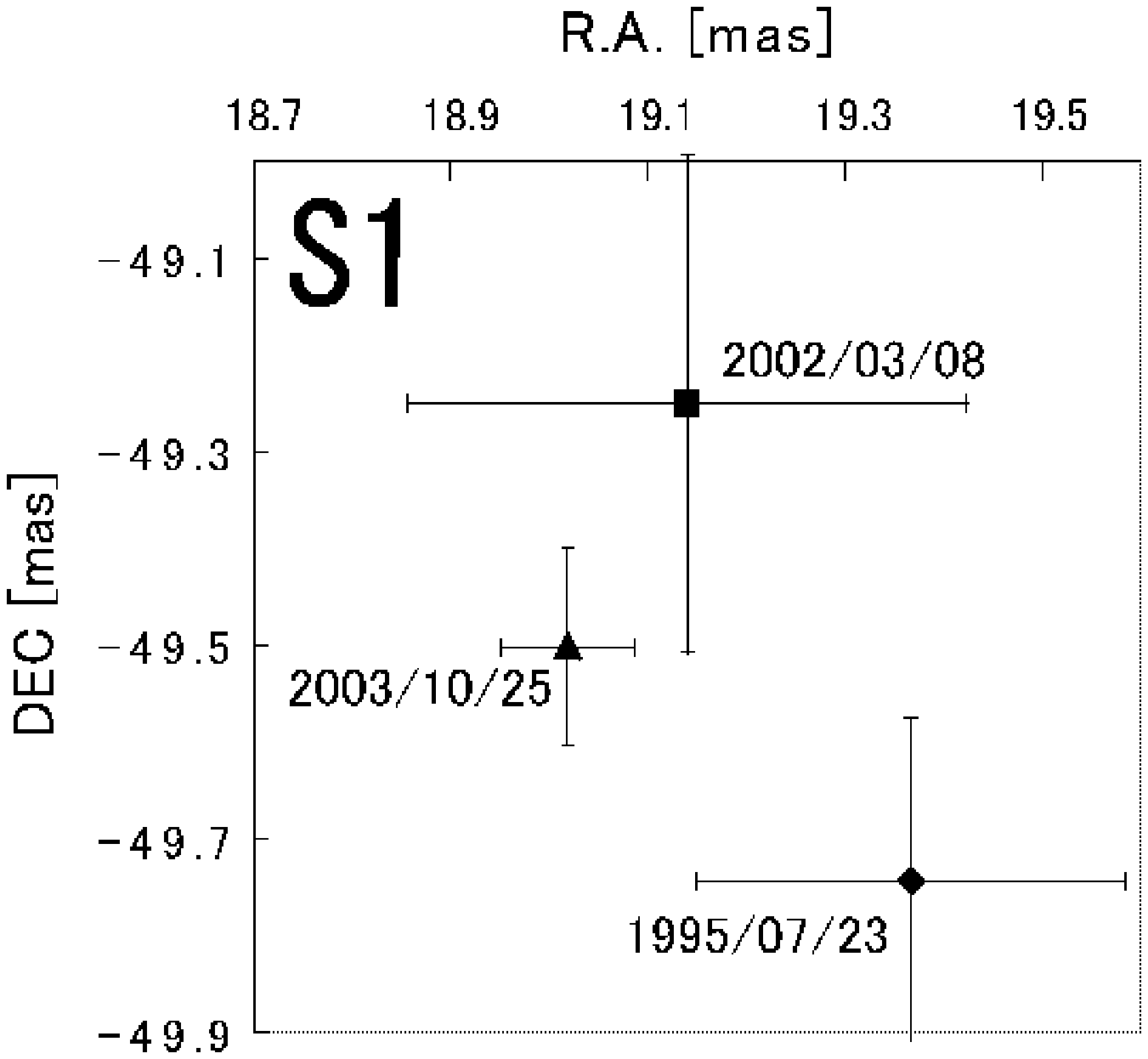}
\end{minipage}
\begin{minipage}{0.33\hsize}
\includegraphics[width=5cm]{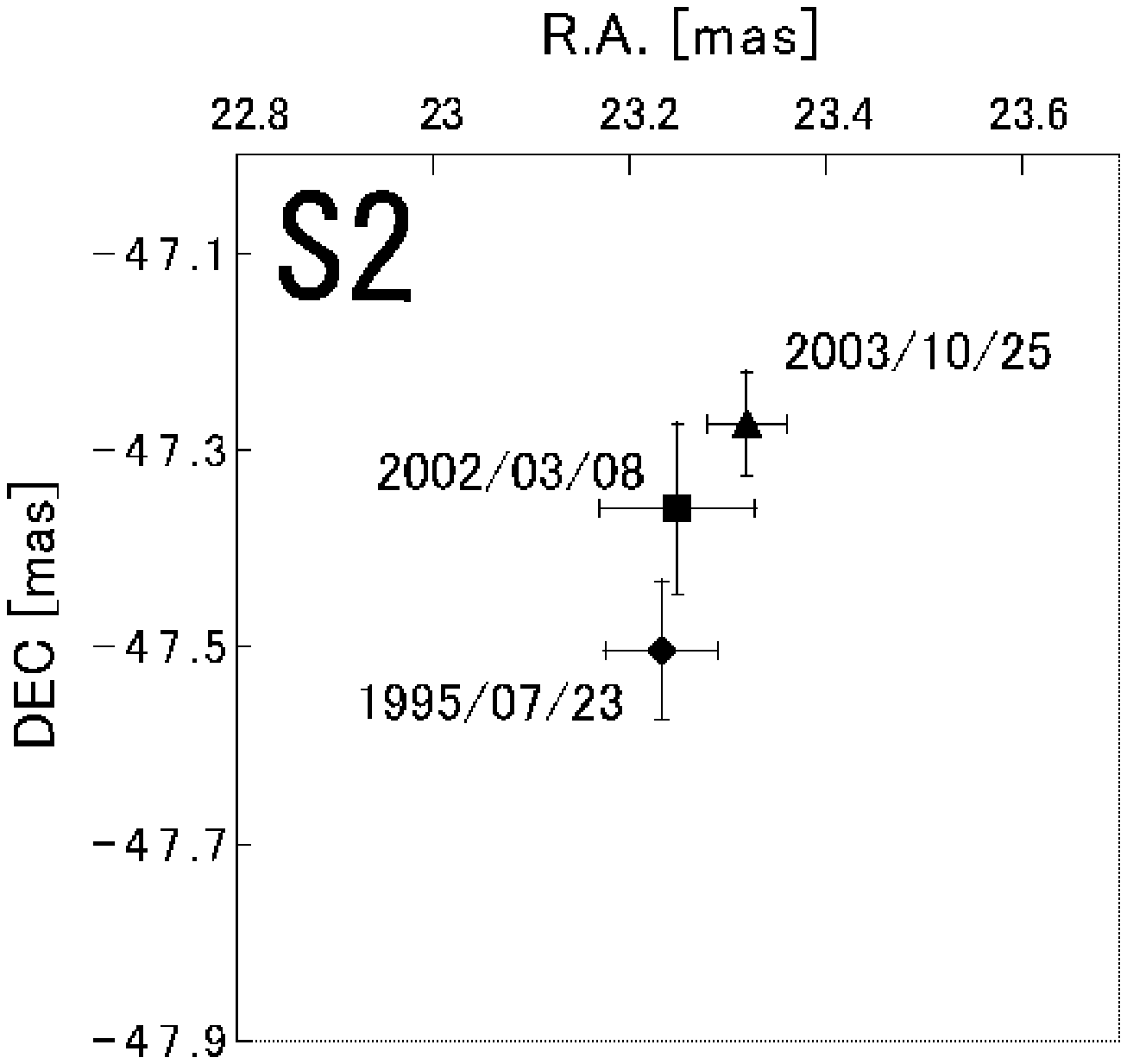}
\end{minipage}
\begin{minipage}{0.33\hsize}
\includegraphics[width=5cm]{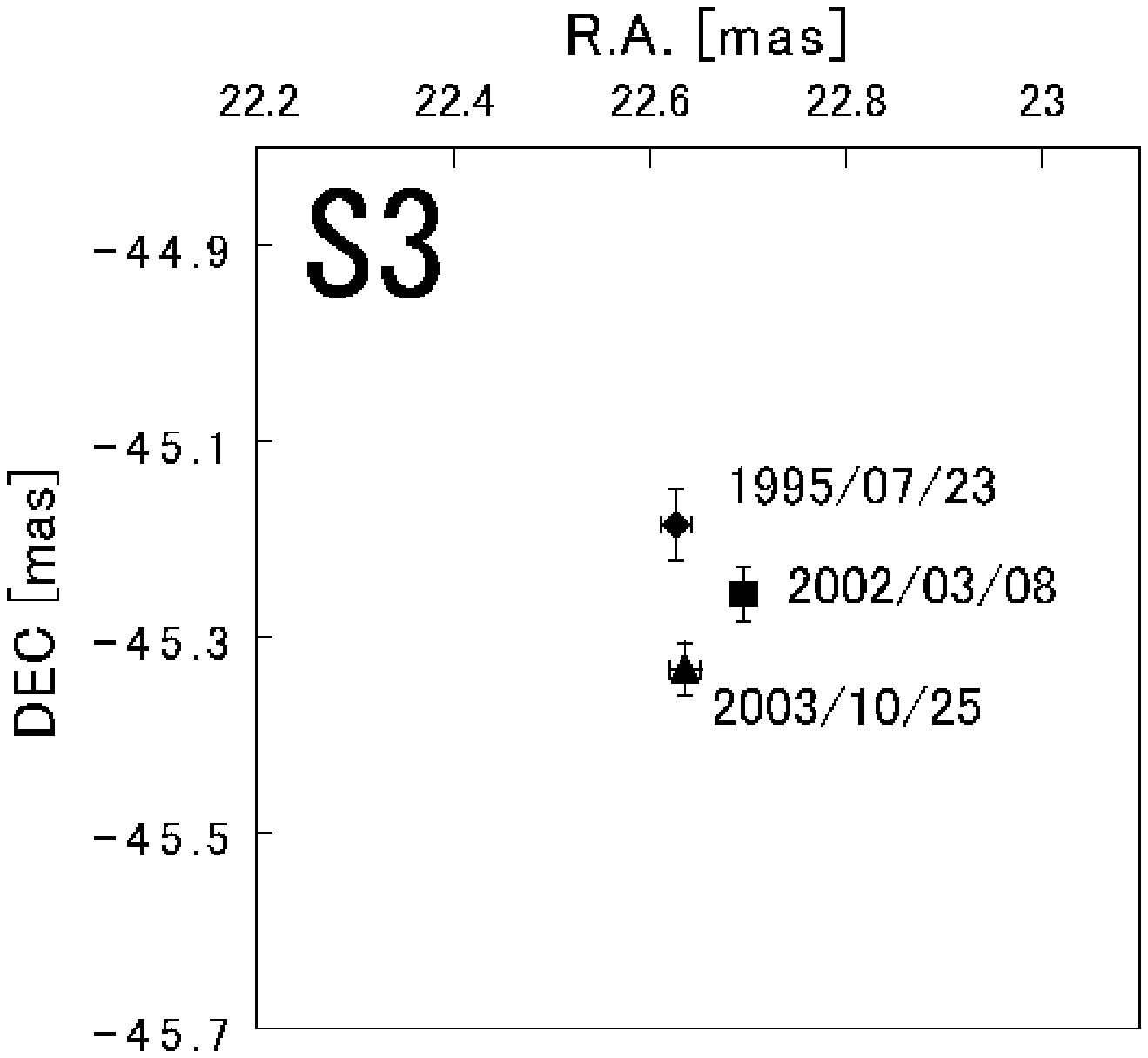}
\end{minipage}\\
\begin{minipage}{0.33\hsize}
\includegraphics[width=5cm]{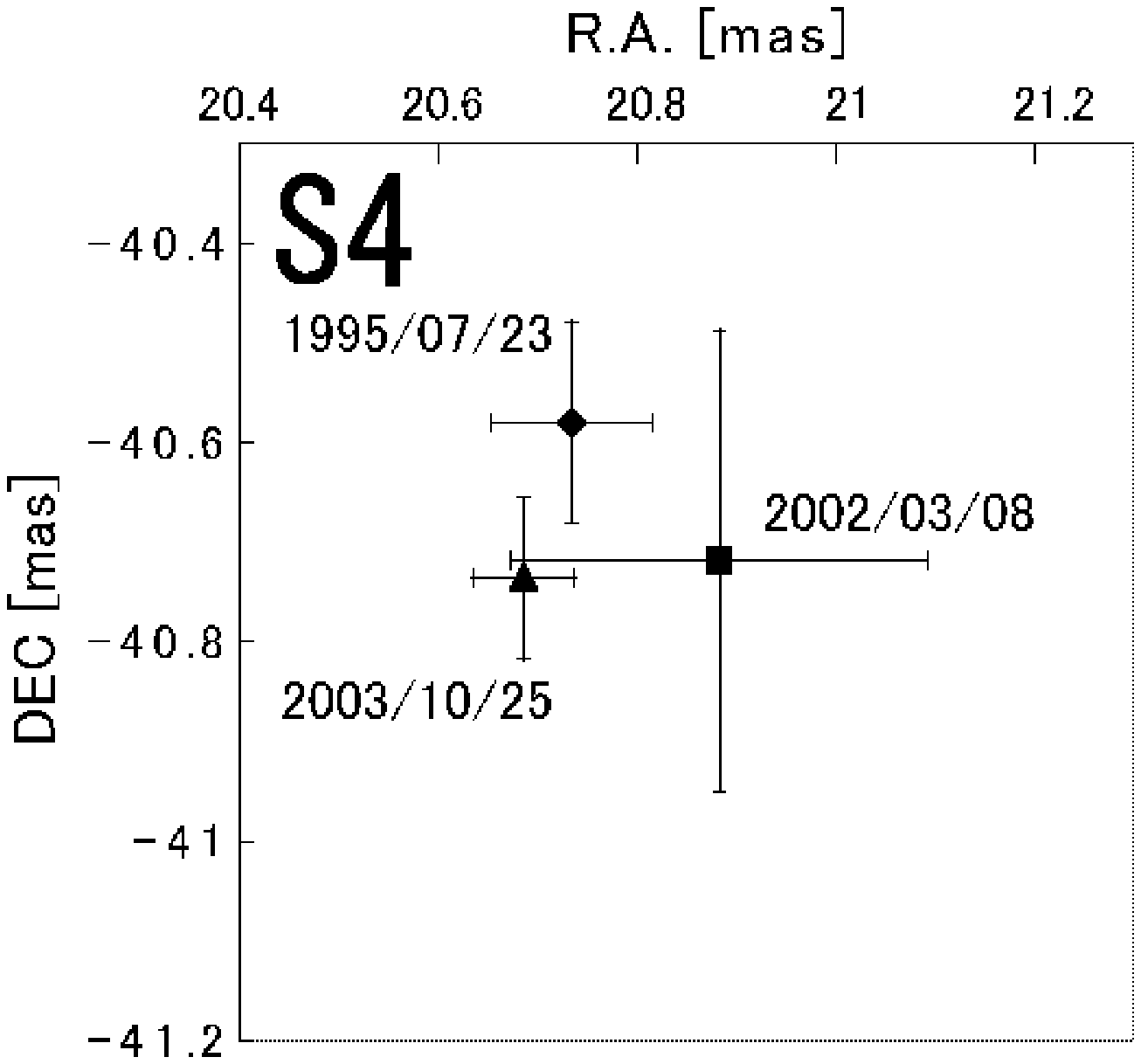}
\end{minipage}
\begin{minipage}{0.33\hsize}
\includegraphics[width=5cm]{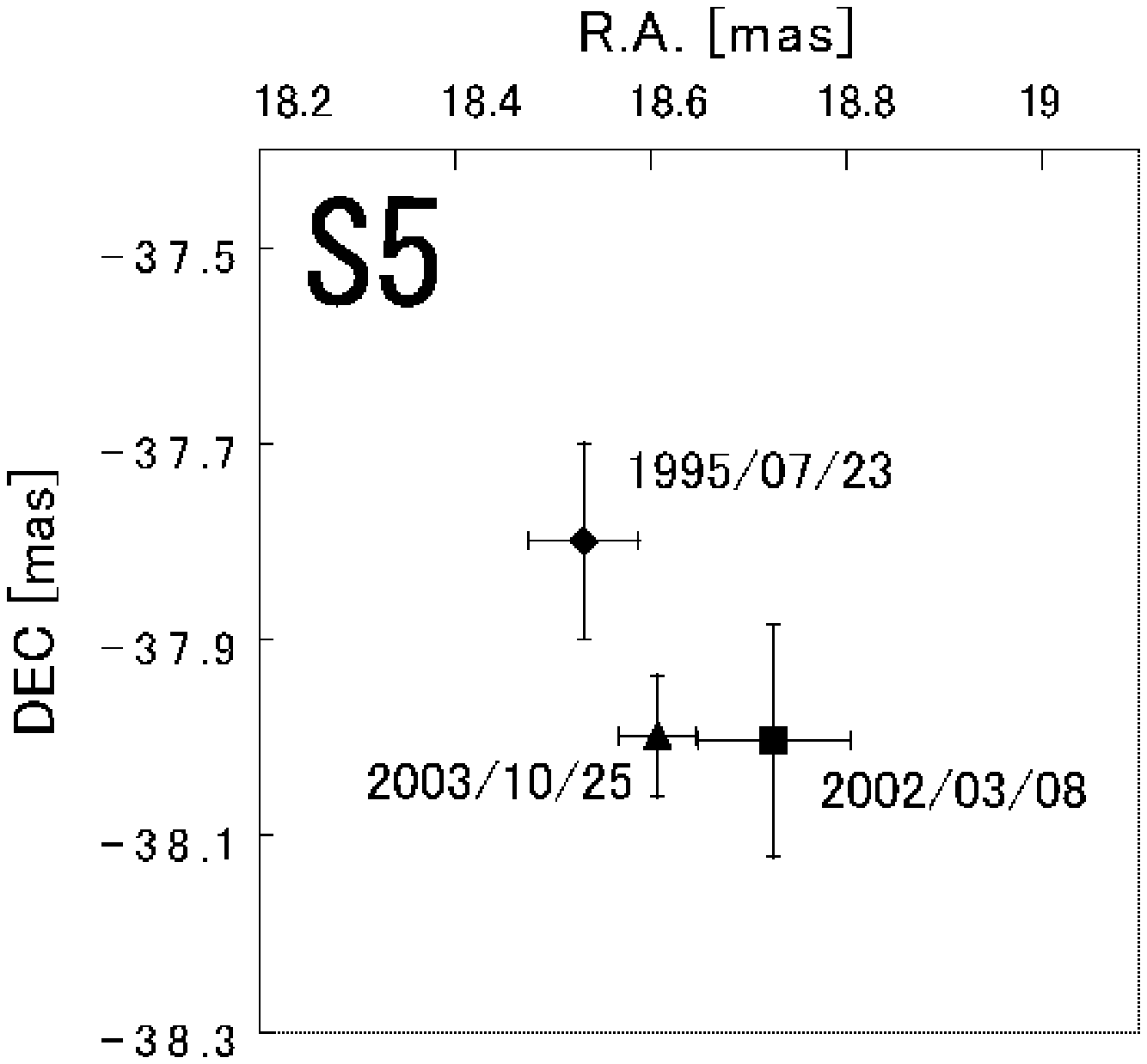}
\end{minipage}
\end{tabular}
\caption{The separation of each component relative to the position of component N1 at 3 epochs.  The error bars are 1~$\sigma$ of Gaussian fits.  The filled diamonds, filled squares, and filled triangles are the results at 1995 July 23, 2002 March 8, and 2003 October 15, respectively.}
\label{motion}
\end{figure}

\begin{figure}
\begin{center}
\plotone{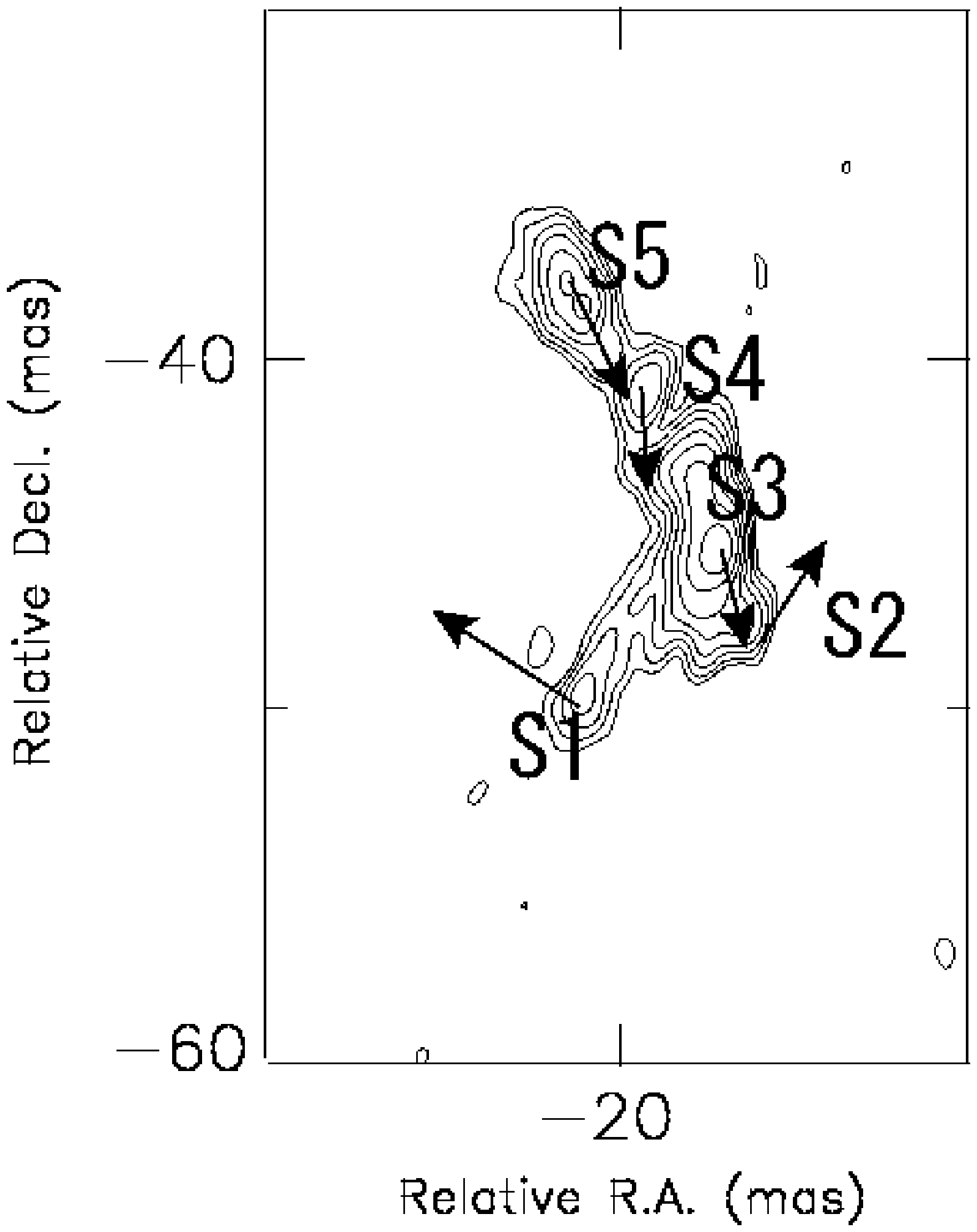}
\end{center}
\caption{The southern lobe of CTD~93 at 15~GHz.  Arrows indicate the direction of motion.  The length of arrow is proportional to the velocity.}
\label{MotionVector}
\end{figure}

We identified the position of each component using Gaussian model fits in the AIPS task JMFIT.  We labeled the components as shown in Figure~\ref{TotalIntensity}(d).  Table~\ref{modelfit} and Figure~\ref{motion} show the position of each component relative to component N1 at the three epochs.  The apparent separation speed of each component is summarized in Table~\ref{velocity}, and the direction of motion is indicated in Figure~\ref{MotionVector}.  We detected a separation change at a significance level of about 3\,$\sigma$ in component S3 and 1.3\,$\sigma$ in component S5.  In contrast, the separation change of component S2 is marginal, and no significant separation change is observed in other components.

\begin{table}[H]
\begin{center}
\caption{The separation of each component relative to the position of component N1.  
}
\begin{tabular}{cccccc} \tableline\tableline
Component & Epoch & $I$ (mJy) & $x$ (mas) & $y$ (mas) & $r$ (mas) \\ \tableline
S1& 1995Jul23 &  7.7 $\pm$ 1.6 & 19.37 $\pm$ 0.22 & $-$49.74 $\pm$ 0.17 & 53.38 $\pm$ 0.27 \\
  & 2002May08 &  9.8 $\pm$ 2.6 & 19.14 $\pm$ 0.28 & $-$49.25 $\pm$ 0.26 & 52.84 $\pm$ 0.38 \\
  & 2003Oct25 & 10.5 $\pm$ 1.2 & 19.02 $\pm$ 0.07 & $-$49.5  $\pm$ 0.1  & 53.03 $\pm$ 0.12 \\ 
S2& 1995Jul23 &  7.7 $\pm$ 1.0 & 23.23 $\pm$ 0.06 & $-$47.5  $\pm$ 0.07	& 52.88 $\pm$ 0.09 \\
  & 2002May08 & 10.5 $\pm$ 1.3 & 23.25 $\pm$ 0.08 & $-$47.36 $\pm$ 0.09 & 52.76 $\pm$ 0.12 \\
  & 2003Oct25 & 13.3 $\pm$ 1.0 & 23.32 $\pm$ 0.04 & $-$47.27 $\pm$ 0.05 & 52.71 $\pm$ 0.07 \\ 
S3& 1995Jul23 & 58.2 $\pm$ 1.8 & 22.63 $\pm$ 0.01 & $-$45.19 $\pm$ 0.04 & 50.53 $\pm$ 0.04 \\
  & 2002May08 & 54.2 $\pm$ 1.5 & 22.7  $\pm$ 0.01 & $-$45.26 $\pm$ 0.03	& 50.63	$\pm$ 0.03 \\
  & 2003Oct25 & 59.8 $\pm$ 1.5 & 22.64 $\pm$ 0.02 & $-$45.33 $\pm$ 0.03 & 50.67 $\pm$ 0.03 \\ 
S4& 1995Jul23 &  6.6 $\pm$ 1.2 & 20.73 $\pm$ 0.08 & $-$40.58 $\pm$ 0.1  & 45.57 $\pm$ 0.13 \\ 
  & 2002May08 &  6.8 $\pm$ 1.8 & 20.88 $\pm$ 0.21 & $-$40.72 $\pm$ 0.23 & 45.76 $\pm$ 0.31 \\ 
  & 2003Oct25 & 10.3 $\pm$ 1.1 & 20.69 $\pm$ 0.05 & $-$40.74 $\pm$ 0.08 & 45.69 $\pm$ 0.10 \\ 
S5& 1995Jul23 & 18.9 $\pm$ 1.7 & 18.53 $\pm$ 0.06 & $-$37.8  $\pm$ 0.1  & 42.1  $\pm$ 0.11 \\
  & 2002May08 & 25.5 $\pm$ 2.2 & 18.73 $\pm$ 0.08 & $-$38.0  $\pm$ 0.12 & 42.37 $\pm$ 0.14 \\
  & 2003Oct25 & 26.5 $\pm$ 1.5 & 18.61 $\pm$ 0.04 & $-$38.0  $\pm$ 0.06 & 42.31 $\pm$ 0.07 \\ \tableline
\end{tabular}
\end{center}
\tablecomments{The flux density of the components is $I$; the separation in right ascension is $x$; 
the separation in declination is $y$; $r=\sqrt{x^{2}+y^{2}}$. }	
\label{modelfit}	
\end{table}

\begin{table}[H]
\begin{center}
\caption{The apparent separation velocity from component N1 between our data and BS025. \label{velocity}}
\begin{tabular}{ccc} \tableline\tableline
Component & Separation velocity ($\mu$as/yr) & Separation velocity ($v/c$) \\ \tableline
S1 &    45.1$\pm$98.0 &    0.71$\pm$1.54  \\
S2 & $-$27.9$\pm$26.1 & $-$0.44$\pm$0.41  \\
S3 &    23.0$\pm$7.7  &    0.36$\pm$0.12  \\
S4 &    23.4$\pm$28.1 &    0.37$\pm$0.44  \\
S5 &    31.7$\pm$22.0 &    0.50$\pm$0.35  \\ \tableline
\end{tabular}
\end{center}
\end{table}

\subsection{Results from Spectral Model Fitting}
\begin{figure}[H]
\begin{center}
\plotone{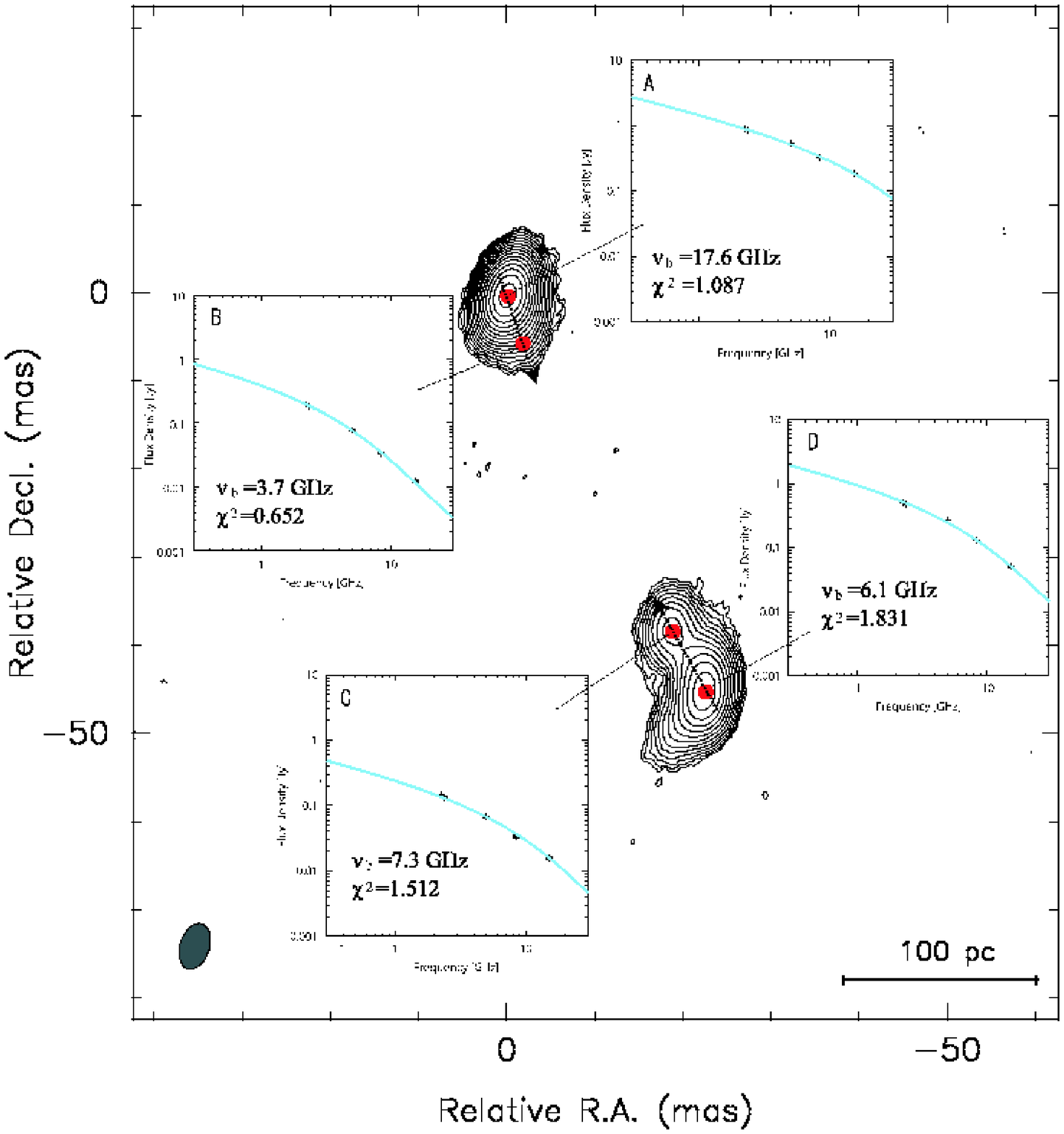}
\caption{Examples of spectral fits for the KP model in the radio lobes.  The contours show the total intensity image at 15.285\,GHz, which is tapered to the same resolution as the 2.244\,GHz image.  The arrows shown in the map are the direction of the profile represented in Figure~\ref{BreakSlice}.}
\label{SpectralAging}
\end{center}
\end{figure}

\begin{figure}[H]
\begin{center}
\plottwo{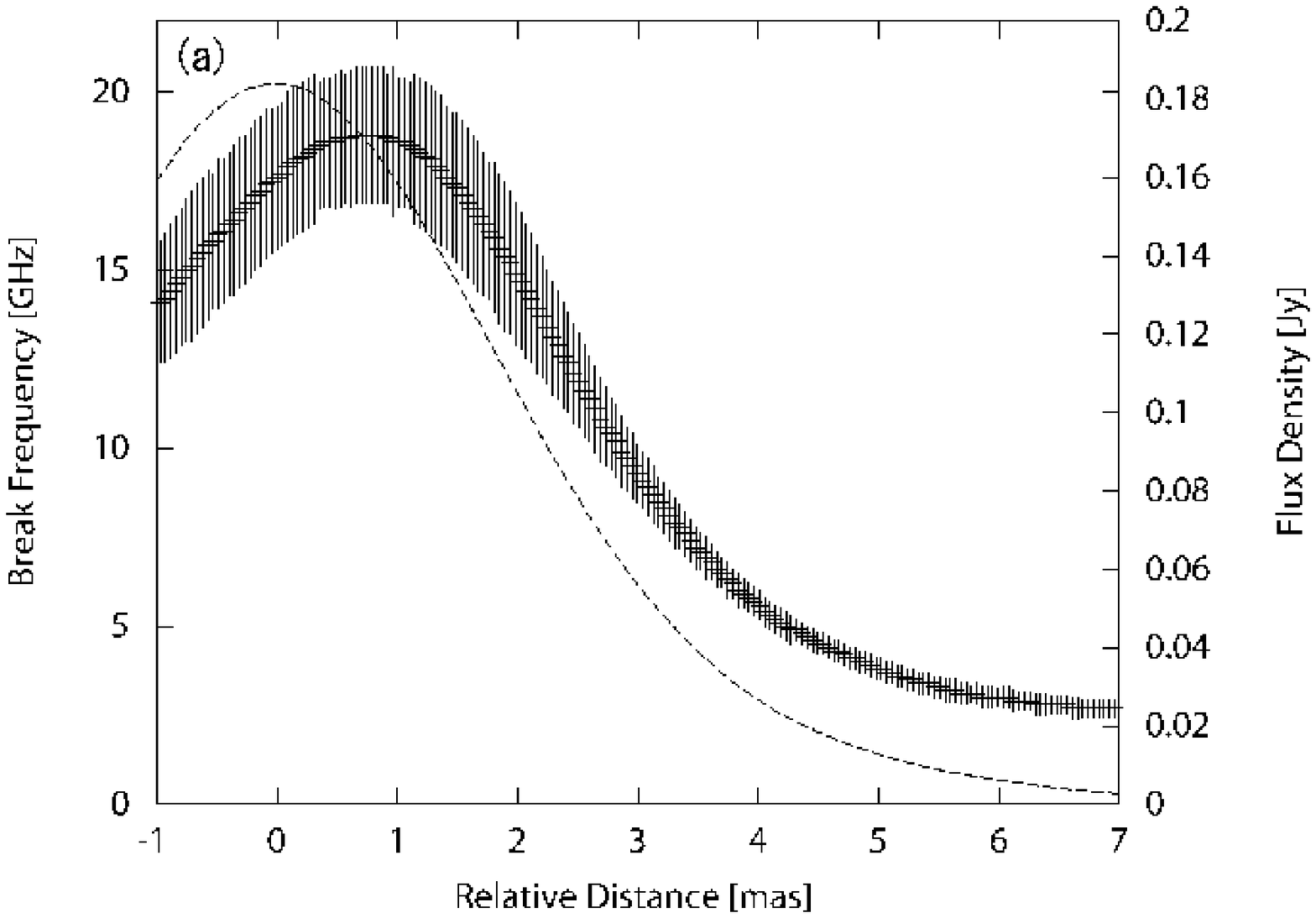}{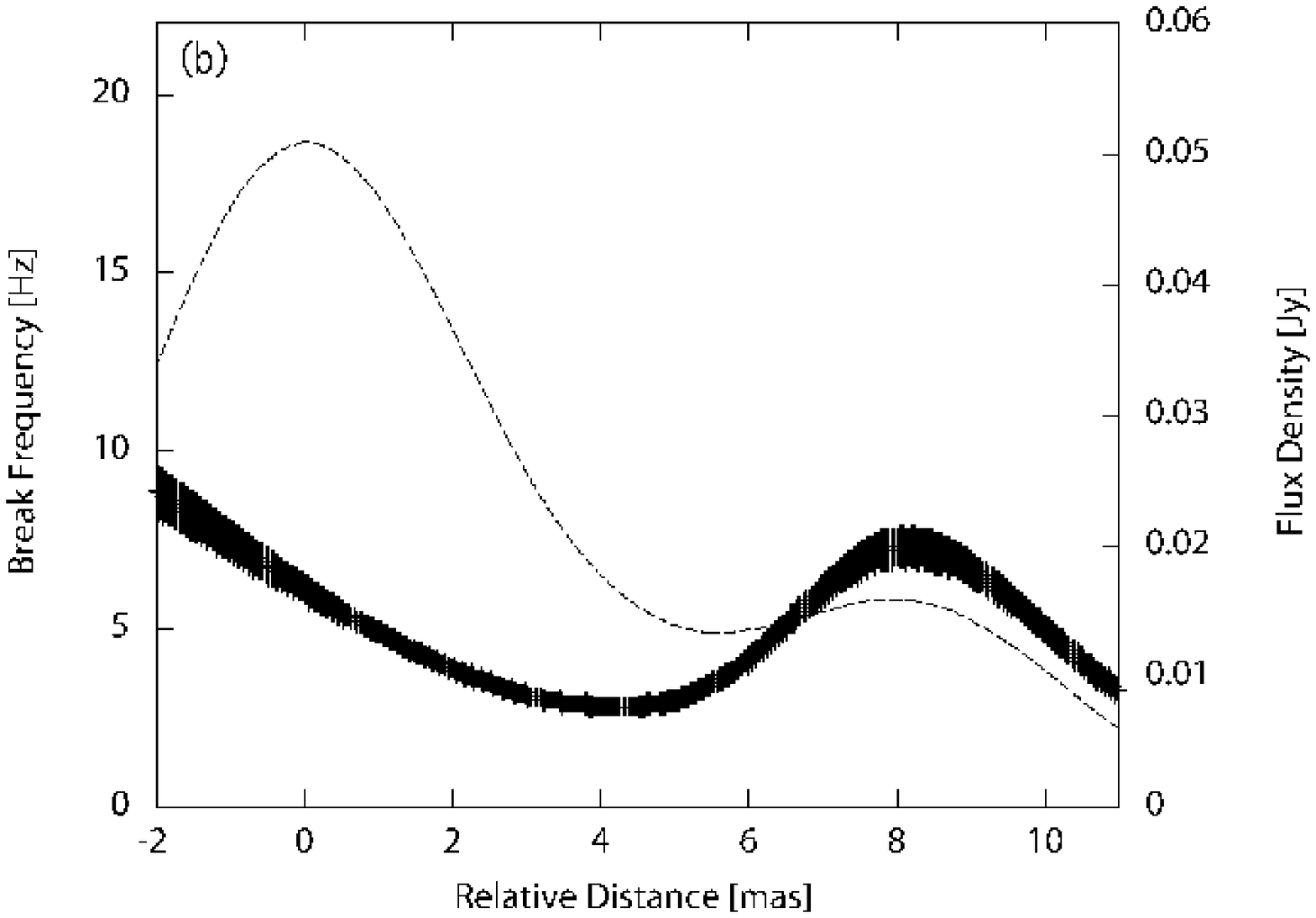}
\end{center}
\caption{The profile of the break frequency (a) in the northern component and (b) the southern component.  The curved lines including error bars show the profile of the break frequency.  The dashed lines show profile of total intensity.}
\label{BreakSlice}	
\end{figure}

Figure \ref{SpectralAging} shows the spectral fitting results for the KP model.  In the fitting process, we first restored all images with the beam size at 2.244\,GHz to match resolutions.  Then the model fitting was performed as described in section~\ref{SpectralModelFitting}.  The inserts in the Figure~ \ref{SpectralAging} show the best-fit spectrum to the observed flux densities at four example points together with the break frequency and reduced $\chi^{2}$ of the fits.  Figure~\ref{BreakSlice} shows the sliced profile of the break frequency along the lines indicated by the arrows in Figure~\ref{SpectralAging}.

\section{DISCUSSIONS}
\subsection{Kinematics}\label{kinematics}
The most significant detection of motion is seen in the hot spot (component S3), which shows a separation rate of 0.36$\pm$0.12$c$.  This separation rate is comparable to that of the hot spots in other several CSOs (e.g., Conway 2002; Gugliucci et al.\ 2005).  The separation speed of this component projected to the source axis is $0.34\pm0.11$$c$.  Assuming the components N1 and S3 are moving apart equal speeds, we derive an advance velocity of 0.17$\pm$0.06$c$.  The resultant kinematic age is $2200\pm700$~yr.  This age is consistent with the youth scenario.  Shaffer et al.\ (1999) attempted to measure the separation rate, but they could not detect significant motion over a period of nearly 20~yr at 18\,cm, probably due to the lower resolution.  Our result is the first detection of the motion in CTD~93.

\subsection{Morphological Interpretation of CTD~93}
Shaffer et al.\ (1999) pointed out that their observations at 2, 3.6, and 6 cm showed characteristics of a core-jet structure with the nucleus located near the northern end of the north component.  They considered that the northern component is the core rather than one of the lobes for the following reasons:\\
1. The wide opening angle of the northern component pointing toward the southwest. \\
2. Faint emission extending to the north of the northern component.\\
3. The prominence of the intensity peak of the northern component at 2 cm.\\
4. The lack of any feature between two dominant components.

If the north component were the core, a counter-jet might be expected to be visible.  The absence of a significant counter-jet component would require relativistic beaming effect due to small angle of approaching jet towards the observer, assuming intrinsic symmetry.  This would imply a relativistic advance velocity for the hot spots.  Another possible explanation is a brightness asymmetry, caused by the variation of external medium density.  In the dentist drill model (Scheuer 1974), such a variation could cause a significant brightness asymmetry between the two hot spots.  It is also possible that there is an intrinsic asymmetry between jet and counter-jet powers.

\begin{figure}
\begin{center}
\plotone{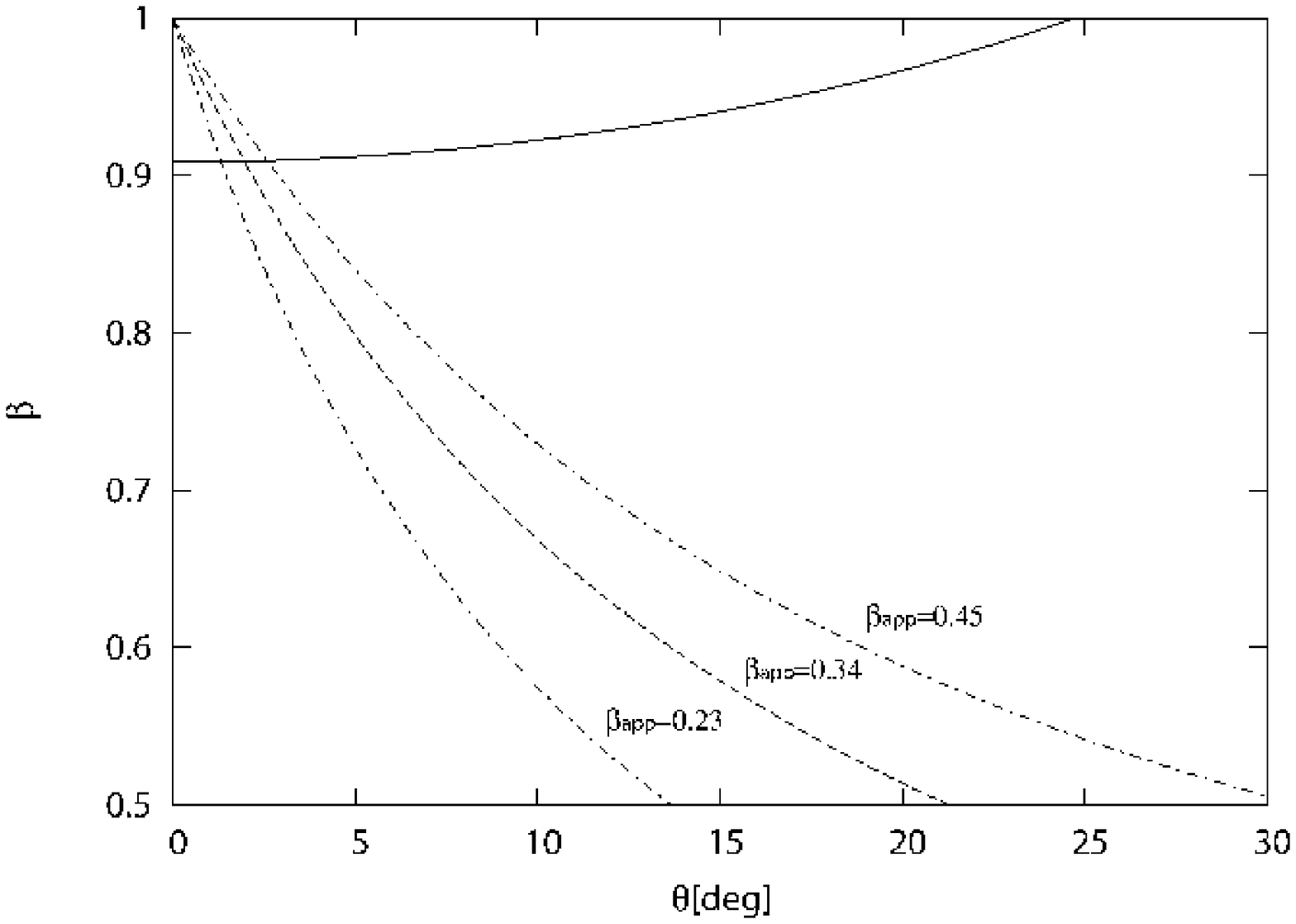}
\end{center}
\caption{The relation between the intrinsic velocity of the hot spot and the viewing angle.  The solid line is $((1+\beta\cos{\theta})/(1-\beta\cos{\theta}))^{2.5}=2000$, and the dashed line is $\beta\sin{\theta}/(1-\beta\cos{\theta})=0.34$.  The dot-dashed lines indicate the 1~$\sigma$ error ranges of the apparent velocity.}
\label{beta-theta}
\end{figure}

Shaffer et al.\ (1999) give the ratio between the peak intensity of southern component and image noise r.m.s.\ as being about 2000 at 5\,GHz.  If the first possibility considered above is the case, the intrinsic velocity of hot spot requires the condition
\begin{equation}
(\frac{1+\beta \cos{\theta}}{1-\beta \cos{\theta}})^{2.5}>2000,
\end{equation}
where $\theta$ is the viewing angle.  Thus the lower limit of the intrinsic velocity is given the solid line in Figure~\ref{beta-theta}.  The intrinsic velocity also requires the relation
\begin{equation}
\beta_{app}=\frac{\beta \sin{\theta}}{1-\beta\cos{\theta}},
\end{equation}
where $\beta_{app}$ is the apparent velocity.  This condition is shown by the dashed line in Figure~\ref{beta-theta} if we assume that the separation rate of the component S3 which we derived in section \ref{kinematics} is the apparent velocity of the southern lobe.  In order to satisfy these two conditions, the intrinsic velocity and the viewing angle must be $\sim0.9$$c$ and less than a few degrees, respectively.  In this case, strong flux variation would be expected to be observed, as often observed in blazars.  According to the long term monitoring observations by the University of Michigan Radio Astronomy Observatory (UMRAO), however, no significant flux variation is observed in CTD~93.  The standard deviation of the flux variation between 1979 and 1999 is about 1\% at 8~GHz.  Thus, the first possibility is excluded.

The other two possibilities are not excluded entirely from our results, but we feel that both possibilities are unlikely.  Many authors have considered the brightness asymmetries of radio lobes and jets (e.g., Fanti et al.\ 1990; Saikia et al.\ 1995; Laing et al.\ 1999; Ishwara-Chandra et al.\ 2001; Saikia et al.\ 2002; P{\'e}rez-Torres \& Breuck 2005).  Although some sources show evidence that brightness asymmetry is not due to the Doppler beaming, the brightness ratios of the two lobes in such sources are significantly smaller than that in CTD~93 ($\ge 2000$) if we assume that CTD~93 is core-jet source.  In addition, the spectral break decreases with distance from the intensity peak of the northern component.  This result is consistent with the basic model for radio lobes involving particle acceleration at the hot spots and with the radio lobes populated by high energy electrons which have leaked out from the hot spots (Blandford \& Rees 1974).  Thus the northern component should be interpreted as one of the radio lobes of a CSO rather than the core of the source.  In following sections, we refer to the northern component as the northern lobe.

In the southern component, the spectral break, similar to the break distribution in the northern lobe, moderately decreases within 5\,mas from the hot spot.  Then a bump of spectral break appears toward northern direction.  The tendency is opposite to that expected from the radio lobe evolution.  In 2-cm image, the southern component is seen as collimated jet and resolved into knots.  This is probably due to the presence of underlying jet.  The location of the bump of spectral break overlaps that of the knot S5 (see the solid line in Figure~\ref{BreakSlice}(b)).  The knots could be produced by the shock in the jet (Blandford \& Konigl 1979) which is causing electron acceleration.  This could be why the break frequency is higher at the knot than at the hot spot.

The break frequency of the southern hot spot is lower than that of the northern one.  Such a difference in break frequency can also be seen in B~1321+321 and B~1943+546 (Murgia 2003).  One possible interpretation for low break frequency of the southern hot spot is that the magnetic field in the southern hot spot is stronger than that in the northern hot spot.  This results in the electrons at the southern hot spot depleting their energy faster than those in the northern hot spot.  Another possibility is the interaction of the hot spot with a dense medium.  Such a dense medium can cause a stronger shock at northern hot spot because the shock strength depends on the density of external medium.  Then the cutoff energy of electrons at the northern hot spot becomes higher than that at the southern hot spot.  A third possibility is that the jets have intrinsically asymmetric power, and the northern jet has generated higher energy electrons.

\subsection{Spectral Age}
We restricted the spectral age calculation in the northern lobe because the break frequency distribution in southern lobe shows complex behavior probably due to the presence of underlying jet.
  To determine the spectral age, the magnetic field strength must be known.  Assuming an ellipsoidal structure of $4\times8$~mas for the northern lobe at 2~cm, we calculate a magnetic field of 15~mG, assuming the standard equipartition condition (Miley 1980).  
It is somewhat uncertain whether the equipartition assumption is adequate.  Recent X-ray observations show that the magnetic field strength in radio lobes is few times less than that of equipartition conditions (Kataoka \& Stawarz 2005; Isobe et al.\ 2005).  Thus our analysis would give a lower limit of the spectral age, which is few to the power of 1.5 (cf., eq.~[3]).  Under the equipartition condition, we derived a spectral age of 300~yr from the equation (3) at the position B ($\sim5$~mas from the hot spot) in Figure~\ref{SpectralAging}.  In the simplified lobe evolution model, the energetic electrons are left behind by the advancing hot spot (Burch 1977, 1979; Winter et al. 1980).  Thus this age must be same as elapsed time since the hot spot passed at the position B.  The implied hot spot advance velocity is 0.26$c$, which shows a good agreement with the hot spot velocity which we derived in section \ref{kinematics}.
		
\subsection{Injection Index Dependence of Fittings}\label{InjectionDependence}
\begin{figure}[H]
\begin{center}
\plotone{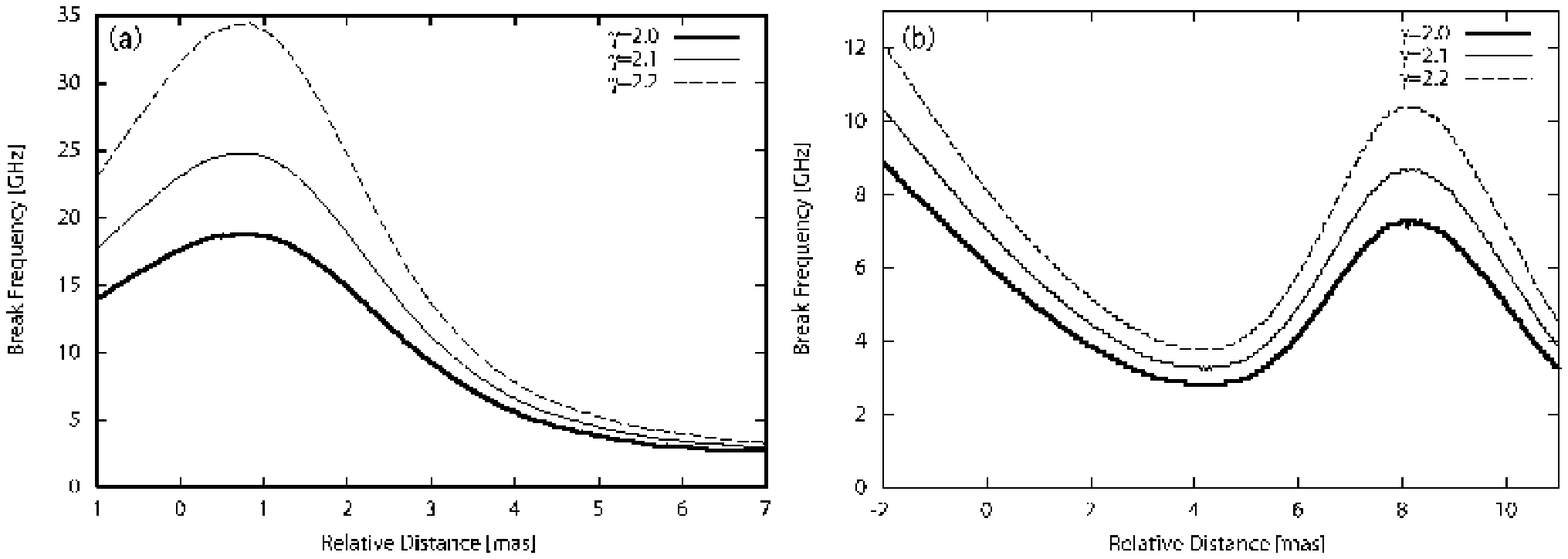}
\end{center}
\caption{The sliced profile of the break frequency in (a) the northern component and (b) southern component.  The thick, thin, and dashed lines indicate the results with $\gamma=2.0$, 2.1, and 2.2, respectively.}
\label{3gamma}
\end{figure}

As we mentioned in section \ref{SpectralModelFitting}, we adopted $\gamma=2.0$ for the spectral fittings.  In this section, we argue how the fitting results are sensitive to the adopted value for the injection index.  To examine the injection index dependence, we fitted with three different values of $\gamma$.  Figure \ref{3gamma} shows the sliced profile of the break frequency for $\gamma$ of 2.0, 2.1, and 2.2.  The break frequency increases with increasing $\gamma$.  For example, break frequencies at position B in Figure \ref{SpectralAging} are 3.7, 4.4, and 5.1 GHz for $\gamma$ of 2.0, 2.1, and 2.2, respectively.  The resultant spectral ages are $\sim$300, 270, 255 yr for $\gamma$ of 2.0, 2.1, and 2.2, respectively.  Although the spectral age varies the injection index, there is a clear trend for decreasing break frequency with distance from the hot spots in any injection index.
	
\subsection{Comparison between KP model and JP model}
As we reviewed in section \ref{sec2}, the spectral aging theory is mainly modeled by two methods, the KP model and JP model.  Even though the JP model is more physically reasonable, the KP model fitted the observed spectrum well in the case of Cygnus~A (Carilli et al.\ 1991).  On the other hand, Murgia (2003) reported that the JP model provides a better fit for some CSS sources.  Tribble (1993) claimed that the JP spectrum in a strong random magnetic field is flattened compared to the standard JP spectrum.  Thus the absence of exponential cutoff of the spectrum might not be a conclusive indication of no pitch angle scattering.  In any case, it is interesting to investigate which of the KP and JP models provides the better fit.

\begin{figure}[H]
\begin{center}
\plotone{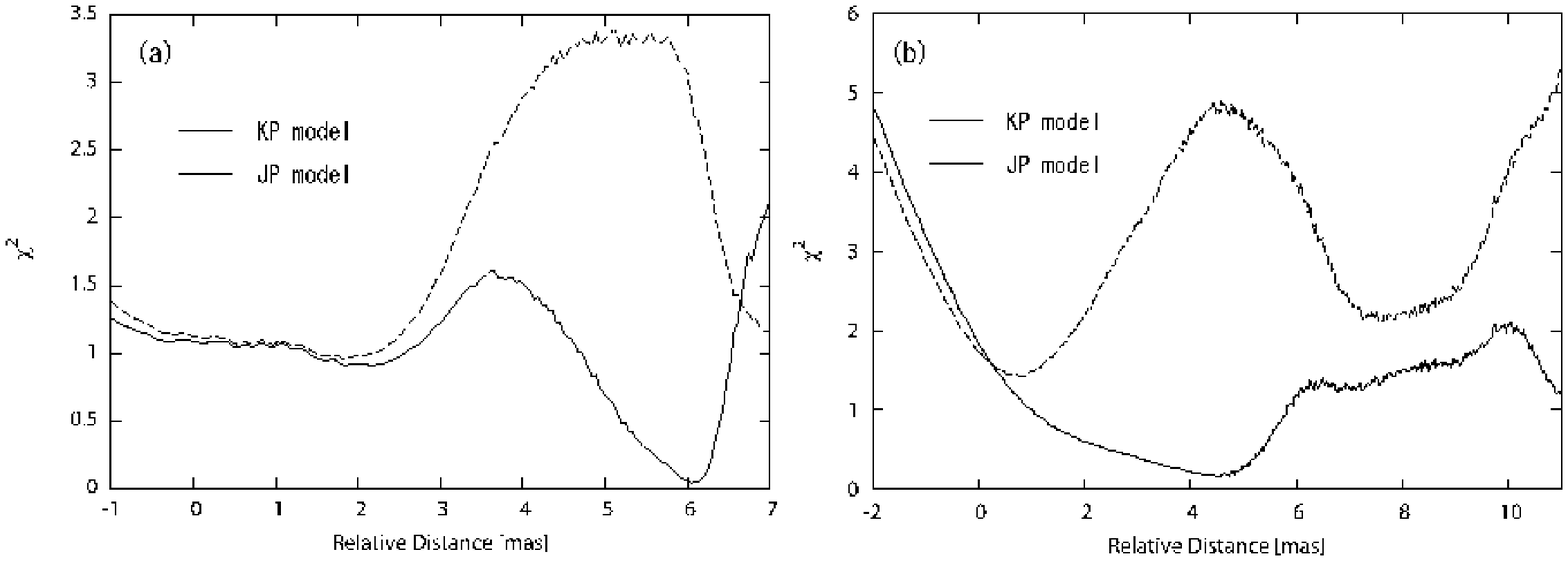}
\end{center}
\caption{Comparison of reduced $\chi^{2}$ between the KP model and JP model in (a) the northern component and (b) the southern component.  Solid lines show the sliced profile of reduced $\chi^{2}$ for the KP model, and dashed lines show that for the JP model.}
\label{KP-JP}
\end{figure}

We have fitted both models to our results.  In Figure~\ref{KP-JP} we show the $\chi^{2}$ distribution of both the KP and JP models along the source axis.  In most regions of the source, the $\chi^{2}$ values of the KP model show relatively small.  The better fits by the KP model are due to a more gradual steepening of observed spectra at high frequency than the exponential cutoff of the JP model.  However, we are aware that this result is not evidence for the KP model.  It can be simply due to the lack of the dynamic range at our observation frequencies.  A wider frequency range data would be required to distinguish between the models.

\section{CONCLUSIONS}
By combining our VLBA observations with archival VLBA data, we have revealed the source expansion of compact radio source CTD~93.  The separation rate between two hot spots along the source axis is 0.34$\pm$0.11$c$, and the corresponding kinematic age is 2200$\pm$700~yr.  This velocity is comparable with the hot spot motion of other CSOs.

Our multi-frequency spectrum measurement of the radio lobes shows high frequency steepening due to synchrotron aging, which is well fitted by the KP model.  Although the KP model shows slightly better fits than the JP model, a wider frequency range observation will be needed to distinguish the KP or JP conclusively.  In the northern lobe, the spectral break decreases with distance from the hot spot.  This is consistent with the basic scenario of radio lobe evolution.  The hot spot advance speed of 0.26$c$, which is derived from the spectral age in equipartition condition, shows a good agreement with the hot spot advance speed that derived from the kinematic aging.

Overall, the results show the northern component is a radio lobe rather than the core/jet as proposed by Shaffer et al.\ (1999).  Thus we favor classification of CTD~93 as a CSO, following Wilkinson et al.\ (1994) and Readhead et al.\ (1996).
	
Our measurements of the kinematic age span a much shorter time than the source age.  In that sense, the hot spot advance we have measured is in fact the instantaneous rate of the separation.  Several mechanisms can cause instantaneous hot spot advance (Owsianik \& Conway 1998), so that the kinematic aging analysis on its own is not strong proof to judge whether the source is youth or frustrated.  On the other hand, the spectral break behavior along the northern component confirms that the source has indeed expanded.  The combination of the kinematic and spectral aging analysis strongly supports the hypothesis that CSOs are young sources.

\bigskip 
We thank P. G. Edwards for reading the manuscript and helpful comments.  We also thank an anonymous referee for constructive comments.
The VLBA is operated by the U.S. National Radio Astronomy
Observatory, a facility of the National Science Foundation (NSF)
operated under cooperative agreement by Associated Universities, Inc.
We have also made use of data from University of Michigan Radio
Astronomy Observatory which has been supported by the University of
Michigan and the NSF.

\end{document}